\documentstyle[aps,preprint]{revtex}
\begin{document}

\title{
NONLINEAR STOCHASTIC DIFFERENTIAL EQUATIONS AND SELF-ORGANIZED CRITICALITY
}

\author{ALBERT DIAZ-GUILERA\footnote{E-mail: albert@ulyses.ub.es}}
\address{
Departament de F\'{\i}sica Fonamental, Universitat de Barcelona \\
Diagonal 647, 08028 Barcelona, Spain
}

\maketitle

\begin{abstract}

     Several nonlinear stochastic differential equations have
been proposed in connection with self-organized critical
phenomena.  Due to the threshold condition involved in its
dynamic evolution an infinite number of nonlinearities arises in
a hydrodynamic description. We study two models with different
noise correlations which make all the nonlinear contribution to
be equally relevant below the upper critical dimension. The
asymptotic values of the critical exponents are estimated from a
systematic expansion in the number of coupling constants by
means of the dynamic renormalization group.

\end{abstract}

\newpage

\section{INTRODUCTION}

     Recently a lot of attention has been paid to the phenomenon
known as self-organized criticality (SOC). Bak, Tang, and
Wiesenfeld (BTW)
\cite{pra38.364} studied a cellular automaton model as a paradigm for
the explanation of two widely occurring phenomena in nature: $1/f$
noise and fractal structures.  Both have in common a lack of
characteristic scales.  This scale invariance suggests that the
system is critical in analogy with classical equilibrium critical
phenomena, but in self-organized criticality one deals with dynamical
nonequilibrium statistical properties.  On the other hand the system
evolves naturally to the critical state without any tuning of
external parameters.

        Several cellular automata models exhibiting self-organized
criticality have been reported in the literature. In the original
sandpile model of Bak. {\em et al.} \cite{pra38.364} the system is
perturbed externally by a random addition of sand grains, once the
slope between two contiguous cells have reached a threshold condition
sand is transferred to its neighbors in a fixed amount.
Taking this model as a reference, different dynamical rules have been
investigated leading to a wide variety of universality classes:
directed flows \cite{pra39.6524}, threshold condition imposed on the
height, on the gradient or even on the laplacian \cite{p179a.249},
continuous variables with a full transfer from a cell instead of a
fixed discrete amount
\cite{prl63.470,p173a.22,pra44.1386,pra45.8551}, deterministic
perturbations in a nonconservative system
\cite{prl66.2669,prl68.2417,pra46.r1720},
anisotropy \cite{pra42.769}, as some examples.

        On the other hand some authors have attempted to connect these
cellular automata models showing SOC to
nonlinear stochastic differential equations
\cite{prl62.1813,prl64.1927,pra45.7002}. These continuous
descriptions are built according to the symmetry rules obeyed by
the
discrete models.
Anomalous diffusion equations with singularities in the
diffusion coefficient have been considered in order to study the
dynamics of the avalanches generated in the critical state
\cite{prl65.2547,prl68.2058}.

        In this paper we study two nonlinear stochastic differential
equations whose deterministic parts keep the symmetries of the
usual sandpile BTW models \cite{pra45.8551,preprint}.
However, they have different noise correlations
from which one gets different critical behavior.
Both models are treated independently and the noise correlations
are obtained by different limits of a general Ornstein-Uhlenbeck
process and in some way they represent different aspects of SOC.
For both models an infinite number of nonlinearities is relevant
below their upper critical dimension $d_c=4,2$; by taking a finite number
of nonlinearities we estimate the values of the critical
exponents from an extrapolation. For one of the models we get
the dynamical exponent obtained in the numerical simulations
of systems exhibiting SOC.

\section{STOCHASTIC DIFFERENTIAL EQUATIONS}

     The nonlinear stochastic differential equation we are
going to study is

\begin{equation}
 \frac{ \partial E(\vec{r},t)}{ \partial t}=D \nabla ^{2}E(\vec{r},t)+
 \sum_{n=1}^ \infty   \lambda _{2n+1}  \nabla ^{2}E^{2n+1}(\vec{r},t)+ \eta
(\vec{r},t)
 \label{series}
\end{equation}
where $D$ is the bare diffusion coefficient and $ \lambda _i$ are the
coupling constants accounting for the nonlinearities of the
model. The variable $E$ which we call energy can have different
physical interpretations \cite{p173a.22}.
The difference among both models lies in the noise correlation

\begin{mathletters}
\label{noise}
\begin{equation}
< \eta (\vec{r},t) \eta (\vec{r'},t')> = 2\frac{ \Gamma }{T} \delta
^d(\vec{r}-\vec{r'}).
\label{quenched}
\end{equation}
\begin{equation}
< \eta (\vec{r},t) \eta (\vec{r'},t')> = 2 \Gamma  \delta (t-t') \delta
^d(\vec{r}-\vec{r'}).
\label{delta}
\end{equation}
\end{mathletters}
$ \Gamma $ being the intensity of the noise and $T$ is a macroscopic
time to be discussed in the last section.

     From Eq. (\ref{series}) we can see that both models are
translational and rotational invariant and that the energy is
conserved in its deterministic part. An additional symmetry ($E
\rightarrow -E$ invariance) has been introduced since in
\cite{preprint} it is showed that it is generated by
a renormalization group procedure when Eq. (\ref{series}) is
derived from an equation involving a threshold condition.

\section{DYNAMIC RENORMALIZATION GROUP}

        The scaling behavior of the system described by the nonlinear
stochastic differential equation is obtained from the correlation function
\begin{equation}
<\left( E(\vec{r}_0,t_0) - E(\vec{r}_0+\vec{r},t_0+t) \right)^{2} >^{
\frac{1}{2} }
 \sim  r^{ \chi } f(t/r^z)
\label{scaling}
\end{equation}
where $ \chi $ and $z$ are the 'roughening exponent' and the dynamical
exponent respectively. The relevance of the different coupling
constants can be checked by naive dimensional analysis: a change of
scale $\vec{r}\rightarrow e^l \vec{r}$ is followed by $t \rightarrow
e^{zl}t$ and $E \rightarrow e^{ \chi l}E$ in order to (\ref{scaling}) be
satisfied. Under this scaling transformation $z$ and $ \chi $ are chosen
to keep the linear equation invariant. With these values one can see
that all coupling constants are of the same order in $ \varepsilon =d_c-d$ and
hence all nonlinear terms are equally relevant for $d<d_c$. The
upper critical dimension $d_c$ are 4 and 2 for models
(\ref{quenched}) and (\ref{delta}), respectively.

     The dynamic-renormalization-group procedure
consists of two steps: an elimination of the faster short wave-length
modes followed by a rescaling of the remaining modes
\cite{pra16.732,pra39.3053}.
The recursion
relations after an infinitesimal transformation for the coefficients
are \cite{preprint}
\begin{mathletters}
\begin{equation}
\frac{dD}{dl}=D \left[ z-2+3\frac{ \Gamma  \Lambda ^{d-4} \lambda _3}{D^3}
 \right]
\end{equation}
\begin{equation}
\frac{d( \Gamma /T)}{dl}=\frac{ \Gamma }{T} \left[ 2z-2 \chi -d \right]
\label{dgdl-q}
\end{equation}
\begin{equation}
\frac{d \lambda _n(l)}{dl}= \lambda _n\left[ (n-1) \chi +z-2) \right]
+ \sum_{j}\gamma_j  \Gamma  \Lambda ^{d-4}D^{a_j}\prod_{i=2}^{n+2} \lambda
_i^{b_{ij}}
\label{el-q}
\end{equation}
\end{mathletters}
and
\begin{mathletters}
\begin{equation}
\frac{dD}{dl}=D \left[ z-2+3\frac{ \Gamma  \Lambda ^{d-2} \lambda _3}{D^2}
 \right]
\end{equation}
\begin{equation}
\frac{d \Gamma }{dl}= \Gamma  \left[ z-2 \chi -d \right]
\label{dgdl-d}
\end{equation}
\begin{equation}
\frac{d \lambda _n(l)}{dl}= \lambda _n\left[ (n-1) \chi +z-2) \right]
+ \sum_{j}\gamma_j  \Gamma  \Lambda ^{d-2}D^{a_j}\prod_{i=2}^{n+2} \lambda
_i^{b_{ij}}
\label{el-d}
\end{equation}
\end{mathletters}
for models
(\ref{quenched}) and (\ref{delta}), respectively.
In the previous equations $\gamma_j$ are numerical
factors and $b_{ij}$ are natural numbers accounting for the vertices
$ \lambda _i$
entering the renormalization of the vertex $ \lambda _n$ in the $j$-th term of
the sum.

Since all terms in this expansion are equally relevant below the
respective upper critical dimension a calculation of the critical
exponents would involve all the contributions. Nevertheless the
exponents can be obtained to first order in $\varepsilon = d_c -
d$ as a series
in the number of coupling constants which are taken into account.
For details of the calculations for model (\ref{quenched}) see
\cite{preprint} where one obtains
\begin{mathletters}
\begin{equation}
\begin{array}{lll}
z=2-0.32(4-d); &  \chi =0.18(4-d) & \mbox{for } d \leq d_c=4 \\
z=2; &  \chi =\frac{4-d}{2} & \mbox{for } d \geq d_c=4
\end{array}
\label{expo-quenched}
\end{equation}
On the other hand the results for model (\ref{delta}) can be
obtained in an equivalent way to be
\begin{equation}
\begin{array}{lll}
z=2-0.33(2-d); &  \chi =0.34(2-d) & \mbox{for } d \leq d_c=2 \\
z=2; &  \chi =\frac{2-d}{2} & \mbox{for } d \geq d_c=2
\end{array}
\label{expo-delta}
\end{equation}
\end{mathletters}

\section{RELATION WITH SELF-ORGANIZED CRITICALITY}

     Up to now it is not clear yet the relation between the
cellular automata models exhibiting self-organized criticality
and the nonlinear stochastic differential equations analyzed in
the present work. In the original sand-pile model, and in others
derived from it, the system is perturbed externally until a site
reaches a threshold condition and then an avalanche is triggered.
During the evolution of the avalanche there is no external
perturbation onto the system. This introduces two time scales in
the model: a fast time scale (the diffusion scale)
and a slow one (the noise scale).

	In order to be suitable for an analytical treatment
one can assume the external noise to be described by a Gaussian
process with zero mean and correlation function given by
\begin{equation}
< \eta (\vec{r},t) \eta (\vec{r'},t')> = \frac{2 \Gamma }{ \tau }e^{-|t-t'|/
\tau }
 \delta ^d(\vec{r}-\vec{r'}).
\label{o-u}
\end{equation}
This is an Ornstein-Uhlenbeck process, with $ \tau $ the correlation
time. This general case can describe two different limits
depending on whether $ \tau $ is a microscopic or a macroscopic time.
In the former case we get (\ref{delta}) whereas in the later one
obtains (\ref{quenched}) with a intensity of the noise that
scales with $1/ \tau $.

     When analyzing the SOC models within a mesoscopic time
scale we deal with the dynamics of single avalanches. In this
case model (\ref{quenched}) is more appropiate since the noise
scale is larger than the time scale we are interested in. A
quenched noise of low intensity does not affect the evolution of
the avalanche since it only depends on whether a site is above
its threshold. This is the reason for which the estimation of
the dynamical exponent (\ref{expo-quenched}) agrees with the
values obtained in the numerical simulations and from scaling
arguments \cite{prl63.470} concerning the evolution of single
avalanches.
Nevertheless it can modify the roughness of the
interface since it can make two distant sites to be correlated
in an artificial way; however, the effect of the nonlinearities
is to lower the value of the roughening exponent with respect
to the linear model.

     SOC models can be analyzed from a different point of view
by considering the avalanches to be instantaneous. In this case
one is mainly interested in the form of the interface after the
relaxation events (avalanches) have taken place and not in its
dynamical evolution. This situation is properly described by
Eqs. (\ref{series}) and (\ref{delta}) but the parameters
entering the deterministic evolution diverge since they are
related to the diffusion coefficient. Only finite values of the
parameters make the problem suitable of an analytical treatment;
on doing so we make the two time scales discussed above to be of the
same order \cite{pra45.7002}.
Although it is not the situation one deals with in
the simulations it is interesting to notice that the roughening
exponent $ \chi $ is lowered with respect to the previously discussed
model. Actually, one should expect a roughening exponent $ \chi  \leq 0$
due to the fact that the energy at each site is bounded after
the avalanches are finished. But as the two dynamics (diffusion
and noise) overlap we can still get a rough interface, at least
for low dimensionality.

     To summarize, we have analyzed two nonlinear stochastic
differential equations by means of the dynamic renormalizaton
group. Both equations share the deterministic evolution but
differ in the noise correlation. Due to the symmetries of the
cellular automata models they are derived from, an infinite
number of relevant nonlinearities must be taken into account. By
taking a finite number of these nonlinear couplings we
extrapolate to infinite and get an estimation of the critical
exponents. These two models permit to describe different, and
complementary,
characteristics
of cellular automata models exhibiting SOC. The quenched noise
model reproduces the dynamical exponent obtained in the
evolution of single avalanches but introduces additional
correlations in the interface profile. On the other hand, the
delta-correlated noise model gives a roughening exponent
which is closer to what one would expect in the profile of the
interface after the avalanches have taken place.

\section*{ACKNOWLEDGMENTS}

This work has been supported by CICyT of the Spanish
Government, grants \#PB89-0233 and \#PB92-0863.

\end{document}